\documentclass[journal]{IEEEtran}
\usepackage{graphicx}
\usepackage{subfigure}
\usepackage{cite}
\usepackage{amssymb, amsmath}
\usepackage{multirow}
\usepackage{epstopdf}
\usepackage{color}
\usepackage{array}

\begin{document}

\title{Edge Learning with Timeliness Constraints: \\Challenges and Solutions}

\author{Yuxuan Sun,~\IEEEmembership{Member,~IEEE,} Wenqi Shi,~\IEEEmembership{Student Member,~IEEE,} 
	Xiufeng Huang,~\IEEEmembership{Student Member,~IEEE,}
	 Sheng~Zhou,~\IEEEmembership{Member,~IEEE,}  Zhisheng~Niu,~\IEEEmembership{Fellow,~IEEE}
\thanks{Yuxuan Sun, Wenqi Shi, Xiufeng Huang, Sheng~Zhou (corresponding author) and Zhisheng~Niu are with Beijing National Research Center for Information Science and Technology, Department of Electronic Engineering, Tsinghua University, Beijing, China.
}}

\maketitle

\begin{abstract}
Future machine learning (ML) powered applications, such as autonomous driving and augmented reality, involve training and inference tasks with timeliness requirements and are communication and computation intensive, which demands for the edge learning framework. The real-time requirements drive us to go beyond accuracy for ML. 
In this article, we introduce the concept of timely edge learning, 
aiming to achieve accurate training and inference while minimizing the communication and computation delay. 
We discuss key challenges and propose corresponding solutions from data, model and resource management perspectives to meet the timeliness requirements.
Particularly, for edge training, we argue that the total training delay rather than rounds should be considered, and propose data or model compression, and joint device scheduling and resource management schemes for both centralized training and federated learning systems. 
For edge inference, we explore the dependency between accuracy and delay for communication and computation, and propose dynamic data compression and flexible pruning schemes.
Two case studies show that the timeliness performances, including the training accuracy under a given delay budget and the completion ratio of inference tasks within deadline, are highly improved with the proposed solutions.

\end{abstract}

\section{Introduction}

\begin{figure*}[!t]
	\centering
	\includegraphics[width=0.95\linewidth]{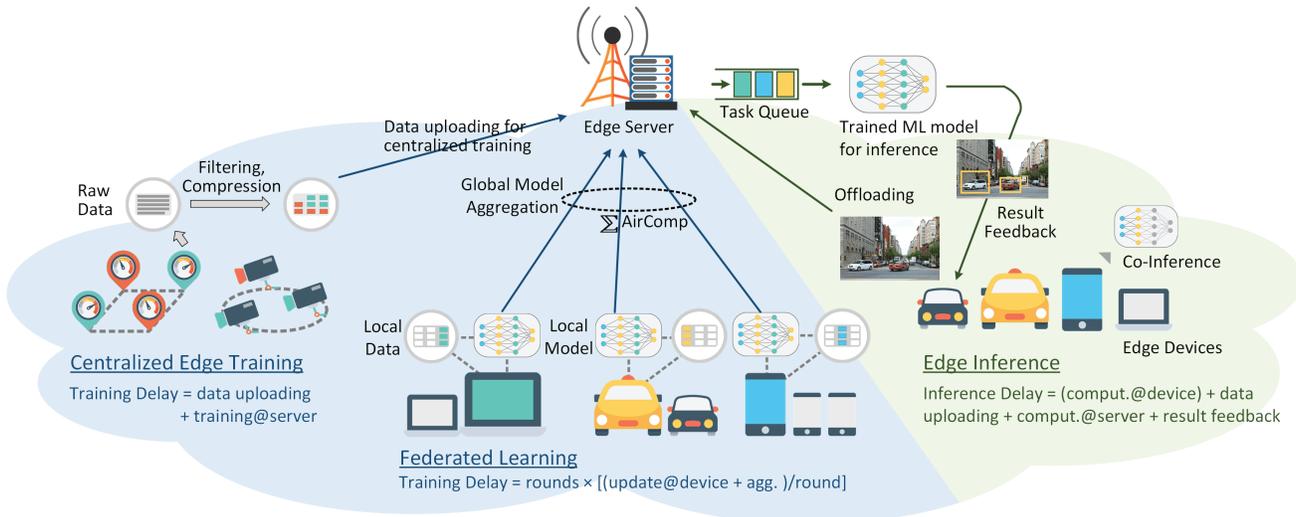}
	\caption{{Illustration of an edge learning system.}}
	\label{system}
\end{figure*}

The convergence of machine learning (ML) and wireless networking facilitates the concept of intelligent edge \cite{Zhu2020toward}.
On the one hand, ML techniques, particularly deep learning (DL), boost the development of many emerging applications, such as autonomous driving, augmented and virtual reality, and industrial Internet. ML also revolutionizes wireless communication technologies from model-based to data-driven, such as channel estimation, auto-encoders, and resource allocation \cite{MLintheair}.
On the other hand, as mobile edge computing (MEC) servers are widely deployed and the edge devices are becoming powerful, the wireless network itself is full of computing capabilities and thus being intelligent. Accordingly, ML models can be trained and implemented at the network edge, namely \emph{edge learning}\cite{Park2019wireless}. Compared with centralized ML in the cloud, edge learning has huge potential to digest the big data, drive \emph{real-time} ML applications and protect privacy. 

An edge learning system includes two key components: \emph{edge training} and \emph{edge inference}.
For edge training, the edge devices can collect or generate real-time data, and upload data to the edge server for centralized training. However, the massive volume of data makes the wireless communication a major bottleneck. Moreover, personal data is exposed to the servers, causing privacy concerns.
An alternative framework is \emph{federated learning} (FL), where edge devices train their local models in a distributed manner based on local data, under the coordination of an edge server who aggregates the global model periodically. 
Meanwhile, it is often difficult to deploy complex ML models such as deep neural networks (DNNs) at edge devices.
Edge inference refers to deploying trained ML models at edge servers, so that devices can offload their inference tasks to edge servers for real-time processing.

Conventionally, the \emph{accuracy} of an ML model is the most important metric. However, many emerging ML applications have real-time service requirements  \cite{6Gwhitepaper}. For example, in autonomous driving, {the delay deadline of inference tasks for environmental perception and object detection is in milliseconds. }
{To enable proactive edge caching for virtual and augmented reality, the time-varying content popularity ranking needs to be learned frequently based on real-time service requirements, to guarantee its freshness and effectiveness \cite{niknam2020federated}.}
These requirements motivate us to \emph{go beyond accuracy}, and introduce \emph{timeliness} as a key factor towards edge learning, with the goal of \emph{fast and accurate} edge training and inference. 
Specifically, timely edge training means that the ML model is trained up-to-date based on the fresh data, so as to adapt to the environmental changes. 
For timely edge inference, tasks should be accurately inferred under stringent delay constraints. 

Although many existing papers on MEC have studied delay-optimized task offloading, they mainly consider general computation models.  
The goal of optimizing the accuracy of ML models motivates ``learning-driven communication" \cite{Zhu2020toward}, where the relation of communication and learning is explored to reduce the \emph{communication} delay for edge training.

In this article, we focus on the timeliness of edge learning specifically for ML tasks.
We jointly consider the two most important yet compromised factors - accuracy and delay, and propose promising solutions from data, model and resource management perspectives, to meet the timeliness requirements for edge training and inference, respectively.
In the following two sections, we first break down the training and inference delay, discuss the challenges and propose solutions for timely edge training and inference, each followed by a case study that validates its performance benefits.
Finally, we conclude the article and discuss some future research directions.

\section{Timely Edge Training} 

As shown in Fig. \ref{system}, a group of edge devices collect or generate real-time data to train an ML model at the edge of wireless network. Each edge device can either transmit its data to the edge server for \emph{centralized edge training}, or employ its local data to jointly train a shared ML model with the coordination of an edge server, namely \emph{FL}.
The two training frameworks are applicable to different scenarios.
For exmaple, centralized edge training can be implemented for Internet of things (IoT) applications, where sensors with very limited computing capabilities are mainly responsible for data collection. Meanwhile, FL is more suitable for privacy-preserved and bandwidth-limited scenarios.

	\subsection{Delay Breakdown and Challenges}
	We define the \emph{training delay} as the time elapsed from data generation to the end of training.
	For centralized edge training, training delay includes \emph{communication delay} for data uploading, and \emph{computation delay} for the edge server to train ML models.
	For FL, training delay is equal to the number of training rounds times the average delay per round, while the latter includes computation delay for gradient update at each edge server, and communication delay for global model aggregation.

	Existing work mainly focuses on the convergence of ML models with respect to the number of \emph{training rounds}.
	{However, the wireless communication resources and computing capabilities of edge devices are limited and time-varying, and thus the communication delay for data or gradient uploading, and the computation delay for model training may vary a lot under different network environments and training policies. Therefore, considering the timely requirements of edge training, we should take \emph{training delay rather than training rounds} into consideration. In particular, the objective is to minimize the total training delay to achieve a certain accuracy threshold, or to optimize the accuracy of an ML model under a given delay budget. }

	There are three key challenges for timely edge training.
	
	\subsubsection{Limited Communication Bandwidth}	
	For centralized edge training, lots of raw data should be uploaded to the edge server in real-time. For FL, the ML model to be jointly learned is of high dimension, and many devices may be involved, requiring high communication bandwidth for periodic global model aggregation.
	The limited communication bandwidth should be efficiently utilized for the most important data samples or model parameters favored by training, 
	in order to reduce the communication delay and speed up the training process.  
	
	\subsubsection{Limited Computing Power of Edge Devices}
	Each edge device has limited yet time-varying computing capability. 
	To update local models for FL, some edge devices may be slower than others and thus become the bottleneck, which is called the \emph{straggler effect}. 
	How to schedule edge devices to address the straggler effect is a challenging issue.

	\subsubsection{Non-i.i.d. Data}
	While data is collected or generated by edge devices in a distributed manner, the local dataset is usually non-independent and identically distributed (i.i.d.), that is, the local data distribution is different from the global data distribution.
	Highly non-i.i.d. data strongly degrades the accuracy and the convergence rate (thus the training delay) of edge training.

	\subsection{Key Solutions}
	We propose some key solutions to address these challenges, 
	aiming at balancing the trade-off between the accuracy of ML models and the training delay, from data, model and resource management perspectives. A summary is given in Table \ref{table1}. 
	
	\begin{table*}[!t]
		\caption{Summary of Possible Solutions for Timely Edge Learning}
		\label{table1}
		\centering
		\begin{tabular}{|m{2cm}<{\centering}|m{3.5cm}<{\centering}|m{1.4cm}<{\centering}|m{7cm}<{\centering}|m{1.5cm}<{\centering}|}
			\hline
			\textbf{Category}&\textbf{Solution}& \textbf{Target} & \textbf{Highlights}  & \textbf{Related Work} \\
			\hline
			\multirow{4}{*}{Edge training} & Data filtering and compression  & data and resource & Filter and compress data samples based on the importance and bandwidth limitation. &\cite{Zhu2020toward}\\
			\cline{2-5}
			& Device scheduling and resource management  & resource  & Balance the trade-off between training rounds and per round communication plus computation delay.&{\cite{Shi2020ICC,yang2019energy,Chen2020Convergence,Liu2020Accelerating,Wang2019Adaptive}}\\
			\cline{2-5}
			& AirComp and gradient compression& model and resource  & Global model is aggregated in an analog way, so that the communication delay does not scale with devices.
			~~ Compress gradients to reduce communication costs.  &\cite{Zhu2020toward, MLintheair,du2020high}\\
			\cline{2-5}
			& {Transfer learning and knowledge distillation} & {model }
			& {ML models are transfered from source to target domains. Use a pre-trained complex model to train a simpler model. }
			&{\cite{Chen2020FedHealth,Jeong2018distillation}}\\
			\hline
			\multirow{2}{*}{Edge inference} & Data compression and task scheduling  & data and resource  & Dynamically compress the input data of inference tasks to balance communication delay and accuracy.& \cite{Huang2020IoTJ}\\
			\cline{2-5}
			& Model compression and resource management  & model and resource  &  {Using pruning or dynamic neural network for model compression, and partition the compressed DNN on servers and devices for co-inference.} & {\cite{cai2020once}}\\
			\hline		
	\end{tabular}	
	\end{table*}

		\emph{1) Data Filtering and Compression:} 

		In centralized edge training systems, the delay to upload a data sample is typically longer than that of training with this sample at the edge server.
		Thus data uploading delay is the major bottleneck under timeliness constraints.
		The key is to make full use of the wireless communication resources to deliver most valuable data samples.		
		
\begin{figure}[!t]
	\centering
	\includegraphics[width=0.9\linewidth]{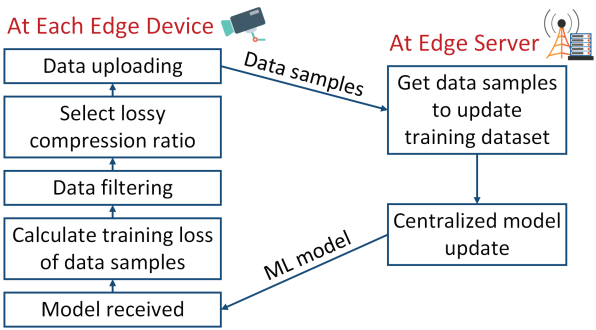}
	\vspace{-3mm}
	\caption{Workflow of data filtering and compression.}
	\label{filtering}
\end{figure}

		Different data samples have different \emph{importances}, that is, the ML model can converge faster when trained with more important data samples. 
		Therefore, spending more communication resources on data samples with higher importance can help the ML model converge faster.	
		Motivated by this, we introduce \emph{training loss} to represent the data importance and guide data uploading.
		As shown in Fig. \ref{filtering}, a training loss based data filtering operation to select important data samples is carried out at each edge device with low processing cost. Given the bandwidth limitation, loss-aware data compression policy then determines the compression ratio of data samples, where less important data samples are transmitted with higher compression ratio and vice versa.
		Data importance can also be characterized by uncertainty, which quantifies how confident each data sample is predicted by the current model  \cite{Zhu2020toward}.
		
		Note that, if the edge server lacks some certain kinds of data, the training loss or uncertainty on these kinds of data samples is typically large. Accordingly, by considering data importance based uploading, the server is also more likely to collect an i.i.d. dataset. 
		
		\emph{2) Device Scheduling and Resource Management:} 
		
		Device scheduling and resource management for FL should consider both the computation delay for gradient update at edge devices, and the communication delay for global model aggregation {\cite{Shi2020ICC,yang2019energy,Chen2020Convergence}}.	
		To attain a certain accuracy, the training delay is equal to the number of training rounds times the average delay per round. 
		On the one hand, scheduling more devices for global model aggregation in each round leads to a longer per round delay, due to the higher probability of having stragglers and less bandwidth allocated to each device. 
		On the other hand, scheduling more devices increases the convergence rate of FL in terms of the number of training rounds, and thus reduces the total number of rounds to attain a certain accuracy. 
		
		By considering the straggler effect of edge devices and managing the communication resources for global model aggregation, the per round training delay is optimized. The relation between total training delay and model accuracy can be derived, which is balanced via device scheduling to meet the timeliness requirements \cite{Shi2020ICC}. 
		{For the unscheduled devices, ML-based estimation techniques can be further leveraged to predict their local model updates at the edge server, in order to improve the convergence rate \cite{Chen2020Convergence}.}
		
		{When communication delay dominates the total training delay while computing resources are relatively sufficient, local update is further enabled in \cite{Wang2019Adaptive,Liu2020Accelerating}, where edge devices implement stochastic gradient descent algorithms for multiple times between two global model aggregations. The key challenge is to optimize the global aggregation frequency to reduce the total training delay \cite{Wang2019Adaptive}, while a momentum term can be further introduced to the local model updates to accelerate the convergence of FL \cite{Liu2020Accelerating}.}

		\emph{3) AirComp and Gradient Compression:}
		
		Many edge devices may be involved for training in an FL system, while ML models often have thousands to millions of parameters. The communication delay of global model aggregation via orthogonal multiple access scales with the number of devices.
		Note that for ML model training, the global model aggregation aims to obtain the \emph{average} of local parameters rather than each individual model. 
		Over-the-air computation, short as AirComp, is proposed as a promising solution \cite{Zhu2020toward, MLintheair}.
		Edge devices simultaneously transmit local models in an analog way, so that all the local parameters are averaged over-the-air.
		By doing this, the communication delay no longer scales with the number of scheduled devices.
		With AirComp, the major trade-off is that, scheduling more devices increases the per round computation delay due to the straggler effect, but reduces the required total training rounds, which should be balanced via device scheduling.
		
		To further reduce the communication delay for model aggregation, local models or gradients can be compressed with \emph{sparsification} or \emph{quantization} \cite{MLintheair, du2020high}. 
		The compression schemes should be designed to keep the key information of the ML model while removing its redundancy.
		Meanwhile, compression ratio is a key design parameter to balance the trade-off between per round communication delay for model aggregation and total training rounds.

		\emph{{4) Other Techniques:}}
		
		{Besides the aforementioned three solutions that aim at balancing the trade-off between the accuracy of ML models and the training delay, there are some other promising techniques for timely training. For example, transfer learning, which transfers the knowledge learned from the source domain to the target domain, can avoid training a model from scratch in the target domain. Knowledge distillation uses a pre-trained teacher model to train a much simpler student model, which can later be implemented at edge devices or servers to meet the timeliness constraints of edge inference. Both techniques can work together with FL \cite{Chen2020FedHealth,Jeong2018distillation} to further improve training efficiency and enable model personalization.}
		
        {
        Finally, we remark that the proposed solutions for FL can be further combined to improve the overall system performance. For example, model compression schemes can collaborate with device scheduling and resource management, and AirComp can also be used in federated transfer learning for global model aggregation.
    }

	\subsection{Case Study: Joint Device Scheduling and Resource Management for Timely Edge Learning}

	We carry out a case study that applies a joint device scheduling and resource management policy for timely edge training \cite{Shi2020ICC}. 
	{A related work on convergence time minimization is done in \cite{Chen2020Convergence}, but the computation delay is not considered.}
	
	Our objective is to maximize the model accuracy within a given total training delay budget. 
	{By considering the straggler effect of edge devices and optimizing the bandwidth resource management among scheduled devices, we quantify how the per round communication plus computation delay grows with the number of devices scheduled in each round.}
	Then based on the convergence analysis proposed in \cite{Shi2020ICC}, we show that the number of training rounds required to attain a certain accuracy decreases with the average number of devices scheduled in each round, and further capture the trade-off between the per round delay and the number of training rounds required. 
	{Our convergence analysis takes the gradient divergence into consideration, so that it can adapt to different data distributions.}
	We formulate an accuracy maximization problem with a training delay budget, and propose a joint device scheduling and resource management policy with fast convergence (FC).
	
	The experiments are performed on an FL system with one edge server and $20$ edge devices. 
	The learning task is the handwritten-digit recognition using the MNIST dataset, with $60,000$ training images and $10,000$ testing images of $10$ digits.
	Both i.i.d. and non-i.i.d. data distributions across edge devices are evaluated.
	In the i.i.d. case, each device is assigned a local training dataset which is uniformly sampled from the whole training dataset.
	For the non-i.i.d. cases, each device randomly selects $1$ or $2$ digits to capture different non-i.i.d. levels.
	A multilayer perceptron (MLP) model with one hidden layer of 64 hidden nodes is trained.
	
	\begin{figure}[!t]
		\centering
		\includegraphics[width=1\linewidth]{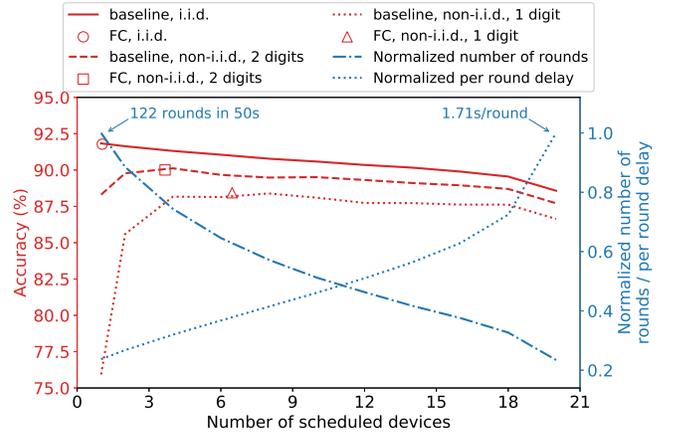}
					\vspace{-5mm}
		\caption{{Performance of the FC policy.}}
		\label{frac-acc}
	\end{figure}
	The experiment results with a total training delay budget $50$s are shown in Fig.~\ref{frac-acc}.
	In the experiment, the average per round delay reaches $1.71$s when scheduling all $20$ devices in each round, while scheduling only one device in each round can have $122$ rounds within the delay budget. Based on these, the normalized number of rounds performed within the training delay budget and the normalized per round delay are shown in the blue curves.
	Results confirm that scheduling more devices in each round increases the per round delay, and thus reduces the possible rounds within the delay budget.
	Each scatter point represents the average number of scheduled devices by the proposed FC policy under a certain data distribution, and the corresponding accuracy.
	Each red curve shows the accuracy achieved by the baseline policy, where a fixed number of devices (reflected by the x-axis) are scheduled in each round.
	We can see that scheduling either too few or too many devices degrades the accuracy of baseline policies, due to the aforementioned trade-off between the required training rounds and the per round delay.

	{The advantages of the proposed policy are as follows. When applying FL in a new environment with unknown data distribution, conventional FL needs to manually tune the number of scheduled devices per round, leading to slow convergence and high tuning cost. Nevertheless, our proposed policy can adapt to various data distributions and resource constraints to achieve fast convergence.}

\section{Timely Edge Inference}

As shown in Fig. \ref{system}, in edge inference systems, the well trained ML models are implemented at edge servers. Edge devices generate tasks and offload them to the edge server for inference. Upon completion, the edge server then sends the inference results back to the device.
Edge devices can also store part of the ML model locally and collaborate with edge server to process tasks, namely co-inference.

\subsection{Delay Breakdown and Challenges}
We define the \emph{inference delay} as the time elapsed from task arrival to the reception of inference results, which mainly includes
\emph{communication delay} for uploading the input data of tasks from device to edge server, and \emph{computation delay} for inference at the edge server.
Since the result is often light-weight, we neglect the result feedback delay.
With co-inference, a \emph{local computation delay} is further introduced before uploading.
Key challenges for timely edge inference are as follows.

\subsubsection{Limited Communication Bandwidth and Computing Capabilities}
The inputs of inference tasks, such as videos and images, are with high data volume. {However, the wireless communication bandwidth is usually limited, making it difficult for multiple devices to upload tasks with low delay.
Meanwhile, the edge server and devices have limited computing capabilities, but the ML-based inference tasks require intensive computations. }
The allocation of communication and computing resources should be jointly optimized to meet the timeliness constraints.

\subsubsection{{Dynamic, Random and Heterogeneous System}}
{	The edge learning system faces many kinds of dynamics, such as the arrival and departure of edge devices due to mobility and the time-varying workloads of edge servers. Wireless channel states and the instantaneous computing speeds often have randomness. To meet the timeliness requirements of inference tasks, it is necessary yet challenging to design online scheduling policies to adapt to the dynamic environment.
Furthermore, inference tasks may have different timeliness requirements, and edge servers and devices are with heterogeneous communication and computing capabilities, which should also be considered for task scheduling and resource management. }

	\subsection{Key Solutions}
	Our solutions aim to reduce the inference delay, by exploring the dependency between inference accuracy and costs of communication and computation. We remove the redundant information of data and ML models, and optimize the compression ratio of data or model and resource management  jointly.

		\emph{1) Joint Data Compression and Task Scheduling:}
		
		The input data of an inference task often has \emph{information redundancy}. By lossy data compression, redundancy can be removed to reduce communication delay. However, higher compression ratio may degrade the inference accuracy. Therefore, the task scheduling scheme needs to dynamically select data compression ratio based on the arrival of tasks and available bandwidth, to balance the trade-off between communication delay and inference accuracy \cite{Huang2020IoTJ}. 
		Task scheduling and compression ratio selection scheme should also consider the computation delay, and make full use of communication resources to maximize the ratio of tasks being inferred successfully under delay constraints.
			
		\emph{2) Joint Model Compression and Resource Management:}

\begin{figure}[!t]
			\centering
			\includegraphics[width=1\linewidth]{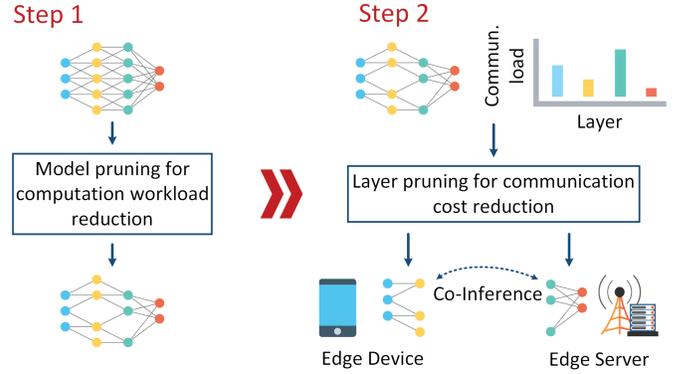}
				\vspace{-3mm}
			\caption{Illustration of a two-step structured pruning.}
			\label{pruning}
		\end{figure}

		To reduce the computation workload of inference tasks, the ML models, particularly DNNs can be compressed.		
		By using structured pruning, some unimportant parts of the original DNN, such as neurons or convolution channels are removed to obtain a smaller, faster and more energy-efficient one, and thus reducing the computation delay.
		The pruned DNN can be further partitioned into two parts and deployed at edge server and device, respectively, for joint inference.		
		Fig. \ref{pruning} illustrates a 2-step structured pruning algorithm. The first step prunes the whole DNN to reduce the total computation workload for inference, and the second step prunes the last layer before offloading to reduce the communication delay for task uploading. Note that, the second step is in fact a generalized data compression method. 
		{The pruned model exported by step 1, and a series of layers, each corresponds to a pruning ratio as the result of step 2, are recorded. These models can be selected on-the-fly with low loading cost according to the available communication and computing resources, in order to enable flexible accuracy-delay trade-off. }
		
		{
		Furthermore, the idea of dynamic neural network can be integrated with resource management \cite{cai2020once}, where a large DNN is first trained and then specific sub-networks can be flexibly selected according to available resources.
		}

		{\emph{3) Other Techniques:}}
		
		{Since the limited computing capability of the edge device is a major bottleneck for timely edge inference, there are many papers focusing on reducing the complexity of the ML models. For example, there are some light-weight architectures such as MobileNet and ShuffleNet. Besides pruning, other model compression techniques such as knowledge distillation and parameter quantization can also be utilized to reduce the computational complexity while avoiding much accuracy degradation.}
		
		{We remark that the proposed solutions above can form flexible combinations. For example, we can implement a compressed ML model while using data compression techniques for task uploading.}

	\subsection{Case Study: Dynamic Data Compression for Timely Edge Inference}
		
	In this case study, we consider delay-constrained edge inference with an edge server and multiple edge devices, focusing on how to reduce the data size for transmission while avoiding accuracy degradation \cite{Huang2020IoTJ}.

	We assume that tasks arrive randomly at each edge device. The edge server schedules devices to send tasks for inference, and sends the results back upon completion. 
	Offloading the input data of tasks is bandwidth and time consuming.
	Each task is given a \emph{delay deadline}, that is, the time interval between task arrival and completion of inference should not exceed this deadline. We focus on communication delay in this case study and ignore the inference time at the edge server. 
	{Data compression overhead is also neglected, but can be incorporated to the communication delay if it is not negligible.}
	A task is successfully completed only when the edge device gets the \emph{correct} inference result within deadline.
	
	To reduce the communication cost for task uploading and complete more tasks before the deadline, lossy compression is used before data transmission. Our objective is to successfully complete more tasks under the constraint of communication bandwidth and delay deadline. A dynamic data compression ratio selection scheme is proposed to balance the trade-off between communication cost and inference accuracy. In the offline case, where the arrival time of each task is known beforehand, dynamic programming is used to design the compression ratio selection algorithm. However, in many scenarios, the system cannot obtain the arrival time a priori. We thus propose an online algorithm based on Markov decision process (MDP). The state is the set of the remaining time for tasks in the queue, the action is the optional compression ratio for transmission and the reward is the accuracy of inference with the selected compression ratio, which can be pre-calculated by performing inference on validation dataset.
	
	Since that the optimal compression ratio of every single task is unknown before being inferred, an \emph{information augmentation} algorithm is proposed. The edge device can first transmit the task with high compression ratio.
	If the result is wrong, the edge server can determine whether to ask the edge device to transmit the task with lower compression ratio. In addition, the correctness of the result may be unknown in real systems. Therefore, the \emph{uncertainty} of the output result is further used to estimate its correctness. 
	Finally, retransmission for combating packet loss due to the unreliable wireless channel is proposed.
	
	\begin{figure}[!t]
		\centering
		\includegraphics[width=1\linewidth]{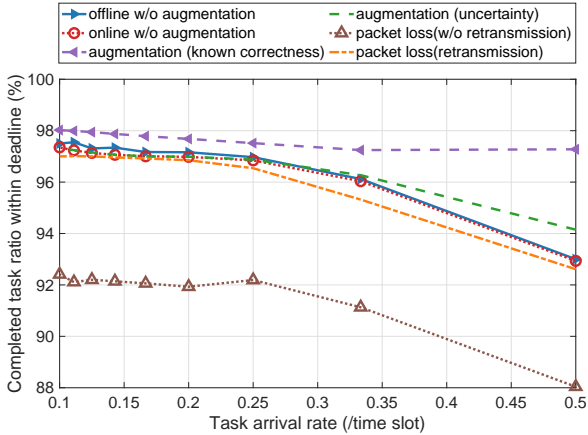}
		\vspace{-5mm}
		\caption{{Task completion ratio of the dynamic compression ratio selection algorithms under different arrival rates.}}
		\label{fig:online}
	\end{figure}

	We use the MNIST dataset to provide inference tasks, and set the deadline of each task to $12$ slots
	{($10$ms for $1$ slot). Each image data sample is with $28\times28$ pixels.}
	Fig. \ref{fig:online} shows the task completion ratio of the proposed algorithms under different arrival rates. 
	Without compression, the ratio of completed tasks is less than $70\%$ when the arrival rate is $0.5$, while the task completion ratios of the proposed algorithms are over $90\%$.
	The information augmentation algorithm brings significant improvement, especially when the arrival rate is high. With unknown correctness of results, the uncertainty-based information augmentation algorithm still performs better than the original online algorithm. 
	When the packet loss probability is set to $0.05$, there is a serious performance degradation without retransmission, but the proposed retransmission algorithm can get almost the same performance as the one without packet loss.

\section{Conclusion and Outlook}
The real-time requirements of ML applications at the network edge call for attentions beyond accuracy.
We therefore introduced the concept of timely edge learning in this article, and proposed promising solutions to achieve accurate and fast training and inference at the network edge.
For edge training, the proposed solutions can balance the trade-off between accuracy and training delay, by the joint data or model compression and resource management.
For edge inference, we explored the dependency between inference accuracy and costs for communication and computation, and optimized the compression ratio of data inputs or ML models based on the availability of computing and communication resources.

As future directions, first, the concept of age of information, which describes the information timeliness and works as a key metric in timely status update systems, can be further used in the edge learning system to describe the timeliness of data, ML model and inference results.
{Second, the energy cost for both edge training and inference is also a key issue. When jointly consider the timeliness requirements and energy budgets, the interplay between delay, energy and accuracy should be further investigated. It is also important to make edge learning greener by reducing the energy costs of edge devices and servers.}
Third, the non-i.i.d. data may significantly degrade the accuracy of ML models and increase the required training rounds, {but the corresponding convergence analysis is at an initial stage.}
To reduce the non-i.i.d. level via data encoding or generative adversarial networks while preserving the privacy of local data is challenging yet important.
{Last but not least, privacy and security issues need to be integrated to the timely edge learning framework, where the reputation of edge devices should be considered to design reliable scheduling schemes.}

\section*{Acknowledgement}
This work is sponsored in part by the National Key R\&D Program of China 2018YFB0105000, Nature Science Foundation of China (No. 61871254, No. 91638204, No. 61861136003), and Hitachi Ltd.

\end{document}